\documentclass[superscriptaddress,twocolumn]{revtex4}
\pdfoutput=1
\usepackage{graphicx,color,url,ulem}
\usepackage{amssymb}

\definecolor{darkgreen}{RGB}{0,139,0}
\definecolor{turqoise}{RGB}{64,224,208}

\definecolor{b}{rgb}{0,0,1.0}
\definecolor{r}{rgb}{1,0,0}
\definecolor{g}{rgb}{0,1,0}

\def\T{\textit{\textbf{T}}}

\begin{document}

\newcommand{\SZFKI}{Institute for Solid State Physics and Optics, Wigner
Research Center for Physics, Hungarian Academy of Sciences, P.O. Box 49, H-1525
Budapest, Hungary}

\newcommand{\BME}{Institute of Physics, Budapest University of Technology
and Economics, H-1111 Budapest, Hungary}

\newcommand{\WAR}{Department of Physics and Centre for Complexity Science,
University of Warwick, Coventry CV4 7AL, UK}

\newcommand{\MDB}{Otto-von-Guericke-University, D-39106 Magdeburg, Germany}

\title{Shear induced alignment and dynamics of elongated granular particles}

\author{Tam\'as B\"orzs\"onyi}

\email{borzsonyi.tamas@wigner.mta.hu}
\affiliation{\SZFKI}
\author{Bal\'azs Szab\'o}
\affiliation{\SZFKI}
\author{Sandra Wegner}
\affiliation{\MDB}
\author{Kirsten Harth}
\affiliation{\MDB}
\author{J\'anos T\"or\"ok}
\affiliation{\BME}
\author{Ell\'ak Somfai}
\affiliation{\WAR}
\author{Tomasz Bien}
\affiliation{\MDB}
\author{Ralf Stannarius}
\affiliation{\MDB}

\begin{abstract}
The alignment, ordering and rotation of elongated granular particles was studied
in shear flow.  The time evolution of the orientation of a
large number of particles was monitored in laboratory experiments by particle
tracking using optical imaging and x-ray computed tomography. The experiments were
complemented by discrete element simulations. The particles develop an orientational order.
In the steady state the time and ensemble averaged direction
of the main axis of the particles encloses a small angle with the streamlines.
This shear alignment angle is independent of the applied shear rate, and
it decreases with increasing grain aspect ratio.
At the grain level the steady state is characterized by a net
rotation of the particles, as dictated by the shear flow.
The distribution of particle rotational velocities was measured both in the steady state
and also during the initial transients.
The average rotation speed of particles with their long axis perpendicular to the
shear alignment angle is larger, while shear aligned particles rotate slower.
The ratio of this fast/slow rotation increases with particle aspect ratio.
During the initial transient starting from an unaligned initial condition,
particles having an orientation just beyond the shear alignment angle rotate
opposite to the direction dictated by the shear flow.
\end{abstract}

\maketitle
\section{Introduction}
\label{intro}
Granular flows involving elongated particles are very important in agriculture,
industry or natural processes. In a shear flow the interparticle contact forces
often lead to orientational ordering and alignment of the particles.
While the basic properties of granular flows have been widely studied for rapid (collisional),
quasistatic and intermediate regimes \cite{beba1994,ehch2003,azra2010,azra2012,mesu2011,boha2005,boec2006,boec2007,boec2009,fehe2003,feme2004,lu2008,brto2011,pega2007,ca2011,jana1996,arts2006,jofo2006,gdrmidi2004,lobo2000,crem2005,sier2001,tari2007,kuma2012},
the flow properties of samples consisting of elongated objects is not well
understood.  After early experiments on
grass seeds \cite{beba1994} and rice grains \cite{ehch2003}, recent numerical
studies focused on shear flows of ellipsoidal particles
\cite{azra2010,azra2012,ca2011}.  These latter studies revealed the topology of
the contact forces \cite{azra2010,azra2012} for sheared ellipsoidal particles
and also described how the stresses and the corresponding particle orientations
depend on the aspect ratio of the ellipsoids \cite{ca2011}.  In a recent work
\cite{bosz2012} we have demonstrated the basic properties of the shear induced
order focusing on the time and ensemble averaged properties only. We
compared these to various systems including colloids, nanoparticles or
nematic liquid crystals. We have also shown that the shear alignment  results
in a considerable reduction of the effective friction of the material.

In the present paper we explore the behavior of individual particles: we
detect their rotation induced by the shear flow, and measure the average
rotation speed as a function of particle orientation during stationary shear
and also during the transient when shear is applied to an initially unoriented
or misoriented system. We also present the basic
properties of the shear aligned state for an extended set of materials.
Details of our experimental and numerical methods are given.

\section{Experimental setup}
\label{exp}

\subsection{Experimental geometry and materials}

\begin{figure} 
\includegraphics[width=\columnwidth]{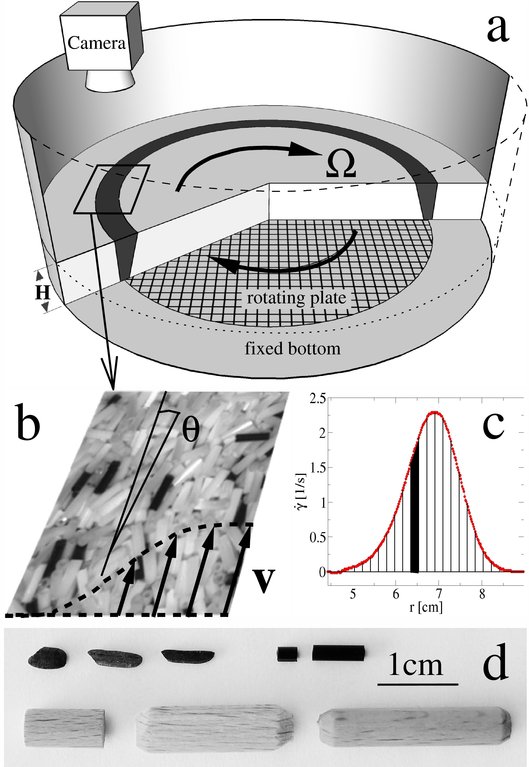}
\caption{(color online). (a) Illustration of the experimental setup.
The granular material consisting of elongated grains is sheared as the circular bottom plate
continuously rotates. Shear strain is localized in the region marked with dark gray.
(b) Example image of the surface of the shear zone for glass cylinders with $d=1.9$ mm and $L/d=3.5$.
The typical orientation of the particles encloses
a small angle $\theta$ with respect to the streamlines, i.e. the particles are 
slightly inclined towards the velocity gradient.
(c) Shear rate $\dot{\gamma}$ as a function of the radius $r$; also
glass cylinders, $L/d=3.5$. Bins with different average shear rate are marked.
(d) Photographs of the particles studied, with $L/d= 2.0$, $3.4$, $4.5$ (rice),
as well as $1.4$ and $3.5$ (glass cylinders), and $2.0$, $3.3$, $5.0$ (pegs).
}
\label{setup}
\end{figure}

The granular material was placed in a so called cylindrical split bottom cell
(see Fig.~\ref{setup}). In this geometry a circular plate at the bottom of the container
is rotated with a
constant speed. If the granular layer is thin
enough (compared to the size of the rotating plate) the middle part of the
granular sample will rotate with the plate like a rigid body. The outer part
of the sample (near the walls of the cylindrical container) remains stationary. The
region in between these undisturbed parts is continuously sheared and is called
the shear zone (or shear band). The shear zone is indicated with dark gray in
Fig.~\ref{setup}(a). Other properties of this configuration, like the precise
geometry of the shear zone and its dependence on the filling height were
described in previous studies \cite{fehe2003,feme2004,diwa2010}.

%

In our experiments, various prolate particles were investigated as shown
in Fig.~\ref{setup}(d).
The length to diameter (or aspect-) ratios of the particles were $L/d$ =
$2.0$, $3.4$ and
$4.5$ (rice grains), $1.4$ and $3.5$ (glass cylinders), and $2.0$, $3.3$, $5.0$
(wooden pegs).  Two experimental methods were used to visualize and
characterize the shear-induced alignment:
(i) complete 3D imaging of the sample with x-ray computed tomography (CT) and
(ii) optical imaging of the particles at the upper surface of the system with a digital camera.
Pegs were investigated by x-ray CT, 
glass cylinders and rice particles by optical imaging. The diameter of the
rotating plate was $D_p\approx 80\, d$, while the container diameter was
$D_c\approx 140\,d$ in both experiments. We used a typical filling height of
$H\approx11d\approx0.14\,D_p$. 

\subsection{Particle detection by x-ray computed tomography}
\label{xray}
The complete 3D arrangement of all particles in the shear zone was determined
experimentally for three samples (pegs with $L/d = 2.0$, $3.3$ and $5.0$) using
x-ray CT \cite{Wegner}.  For these measurements we used a medical angiography system,
i.e., a rotational c-arm based x-ray device equipped with a flat-panel detector
\footnote{The x-ray measurements took place at the INKA-Lab of the Otto
von Guericke University, Magdeburg (http://www.inka-md.de).}.
The spatial resolution of the device was between
1.5~pixel/mm and 2 pixel/mm.  The primary purpose of these experiments was to
show that the order and alignment data extracted from surface particles are
representative for the global order in the shear zone, with only slight
variations.  The particles were detected by applying a watershed algorithm to
the recorded images.
The particles' orientation was defined as the direction of the largest
eigenvector of the individual moment of inertia tensors.
The shear-induced orientational order is monitored by diagonalizing the
symmetric traceless order tensor \T{}:
\begin{equation}
T_{ij}= \frac{3}{2N} \sum\limits_{n=1}^N \left[{\ell}^{(n)}_i {\ell}^{(n)}_j
-\frac{1}{3} \delta_{ij}
\right] \quad ,
\end{equation}
where $\vec {\ell}^{(n)}$ is a unit vector along the long axis of particle
$n$, and the sum is over all $N$ detected particles.
The largest eigenvalue of \T{} is the primary order parameter $S$.
The shear-induced alignment is characterized by the average alignment angle
$\theta_{\rm av}$, which is the angle of the corresponding principal axis of \T{}
measured with respect to the flow direction [see Fig.~\ref{setup}(b)].
Positive $\theta$ corresponds to inclination towards the velocity gradient.
 The components of \T{} were determined
both near the surface (particles with centers in the top $\approx 1$~cm
layer, approx.\ 2 particle diameters) and in the bulk (2~cm thick layer below),
corresponding to about 10,000 and 20,000 particle positions, respectively.
The experiment is described in detail in \cite{Wegner}.

\subsection{Optical particle detection}
\label{optic}

For rice grains and glass cylinders, the top surface of the whole sample was
monitored with a digital camera using a frame rate of 18 fps. As an illustration,
a section of an image is shown in Fig.~\ref{setup}(b). The shear rate $\dot{\gamma}$
is not constant across the shear zone, but it changes with the radius $r$
measured from the center of the cell. It is expressed as
$\dot{\gamma} = {\rm d} v(r) / {\rm d} r - v(r)/r$, where $v(r)$ is the tangential
velocity of the material. The tangential velocity $v(r)$ was determined using a
self developed PIV (Particle Image Velocimetry) code and the time averaged
shear rate was calculated. This shear rate profile is presented in
Fig.~\ref{setup}(c).  The shear zone was divided into about 20 bands (widths
0.19 cm $\approx d$) as illustrated in Fig.~\ref{setup}(c). The average
shear rate was determined for each band.  The orientation and in-plane length
of colored tracer grains (6$\%$ of the sample) were detected, 
yielding about 60,000 particle positions in each band for each type
of grain.  This way, a single experiment with constant rotation speed $\Omega$
provides information about the shear rate dependence of the alignment \cite{lu2008}.
In order to cover a wide range of shear rates, the experiments were repeated
at three different values of $\Omega$.

\subsection{Torque measurement}
\label{torque}

In order to monitor changes in the mechanical resistance of the sample against
shearing, we continuously measured the torque needed to maintain a constant rotation
rate of the bottom plate during the evolution of the alignment. This has been done
the following way: The driving motor was mounted on a rotating table. 
To keep the body of the motor fixed and let the plate rotate in the sample container, 
we use a load cell attached to the body of the motor.
We measure the torque needed to hold the motor while the granular material
is sheared with a constant speed by the rotating bottom plate.

\section{Numerical setup}
\label{num}
Discrete Element Method (DEM) simulations were performed to complement the experiments.
A modified version \cite{brto2011} of the general purpose molecular dynamics code
LAMMPS \cite{pl1995} was used to generate stiff elongated particles by gluing spheres
together. Long particles generated such way have parts with concave curvature and
thus are prone to interlock if their structure is identical. Therefore we prepared
polydisperse samples by allowing the variation of the following quantities: the
radius of the spheres by 10\%, the length of the particles by 10\%, and the overlap of the
spheres between 30-70\% of the radius. We also varied the number of spheres required to
build up a long particle.
%
\begin{figure}[ht]
\includegraphics[width=\columnwidth]{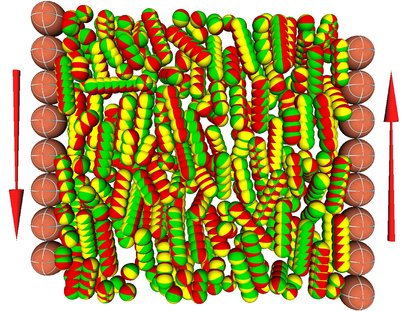}
\caption{(color online). Example cut of the numerical setup. The wall consisting of
larger spheres imposes constant pressure and stationary shear to the sample.
The snapshot was taken in the steady state, where most of the particles are shear aligned.
}
  \label{numdemo}
\end{figure}

The dynamics of 700-1200 elongated particles was modeled in simple
Couette flow, where the parallel walls were moved in the opposite directions
as indicated in Fig. \ref{numdemo} by the arrows. The walls consisted of
twice as large spheres as the average particle diameters.
In the other two directions periodic boundary conditions were chosen. 
The particles were confined with constant pressure during shear.
The generated shear rate was relatively uniform across the sample
\cite{shbr2012}.

\section{Results}
\label{results}

\subsection{Stationary state}

\begin{figure}[ht]
\includegraphics[width=\columnwidth]{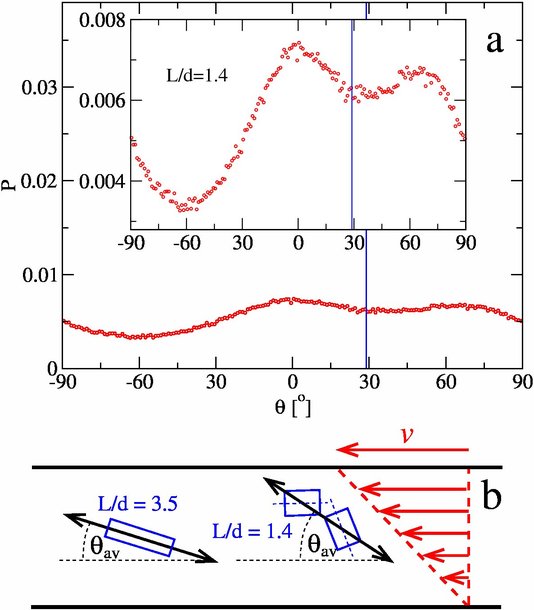}
\caption{(color online). (a) Distribution of the orientational angle
$\theta$ of the particles with respect to the streamlines for cylinders with
$L/d = 1.4$. Data obtained by optical particle detection.
The inset in panel (d) is the same data on a different scale.
The vertical blue line corresponds to the average orientation,
$\Theta_{\rm av}$.
(b) Illustration of typical (average) particle alignment for longer ($L/d =3.5$)
and shorter $L/d = 1.4$ cylinders.
}
  \label{dist1_4}
\end{figure}

As described in Section \ref{exp}, a sufficiently large set of particle positions and 
corresponding orientations were detected using either optical imaging at the surface
or x-ray CT in the whole sample.
%
\begin{figure}[ht]
\includegraphics[width=\columnwidth]{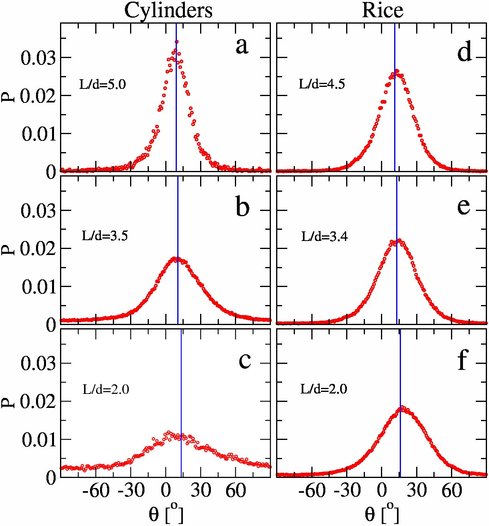}
\caption{(color online). Distributions of the orientational angle
$\theta$ of the particles with respect to the streamlines for (a)-(c) cylinders with
$L/d = 5.0$, $3.5$, and $2.0$, and (d)-(f) rice particles with $L/d = 4.5$, $3.4$,
and $2.0$, respectively. For panels (a) and (c) data obtained by x-ray CT, 
other data obtained by optical particle detection. The vertical blue line corresponds to the stationary average
orientation, $\Theta_{\rm av}$.
}
  \label{dist}
\end{figure}
The distribution of the alignment angles $\theta$ (measured in the shear
plane with respect to the local direction of the streamlines) is presented
for cylinders with length to diameter ratios of $L/d = 5.0$, $3.5$,
and $2.0$ and and rice with  $L/d = 4.5$,  $3.4$,
and $2.0$ in Figs.~\ref{dist}(a)-(c) and Figs~\ref{dist}(d)-(f), respectively.
The first thing to note is the aspect ratio
dependence of the distributions. The graphs show that increasing $L/d$ leads
to a narrower distribution, i.e. a higher degree of shear induced ordering.
The time and ensemble averaged particle orientation $\Theta_{\rm av}$ was determined
by fitting the data by a Gaussian (with the constant offset subtracted), 
and is illustrated with a vertical (blue/gray) line in  Fig.~\ref{dist}(a)-(f).
Thus, asymptotically the average orientation encloses a small angle $\Theta_{\rm av}$ with
the streamlines, as it is illustrated in Fig.~\ref{setup}(b) for $L/d =3.5$.
The second interesting observation is that for cylinders ($L/d =
3.5$ and $2.0$) the distribution is broader than for ellipses with similar aspect ratio.
This can be understood by analyzing the shortest sample ($L/d = 1.4$) 
in Fig.~\ref{dist1_4}. Here, the distribution has two peaks, see the
zoomed inset of Fig.~\ref{dist1_4}(a).
%
This suggests that these grains align with their longest extension (diagonal) in
the preferential direction. Since we determine the particle orientation by
measuring the orientation of the symmetry axis, the distribution of the detected
angles has two peaks, approx.\ $2\arctan(1/1.4)=71^\circ$ apart, which correspond
to the two particle configurations illustrated in Fig.~\ref{dist1_4}(b) for $L/d =1.4$.
The effective average alignment angle in this case is between the angles corresponding
to the two peaks, i.e.  around $\Theta_{\rm av}=29^\circ$.
Getting back to the case of longer cylinders, the above described effect could be one of 
the factors leading to 
a widening of the distributions (especially for the case of cylinders with sharp edges)
compared to the case of rice with similar aspect ratio
(compare Figs.~\ref{dist}(c) and \ref{dist}(f)).

Figure~\ref{dist} also illustrates the typical scatter of the angular
distributions obtained by the two methods (optical or x-ray measurements).
Data for all rice samples as well as for cylinders with $L/d = 3.5$ were obtained by
optical detection and the histograms presented were created using about 200,000
data points, while the data for pegs with $L/d = 5.0$ and
$2.0$ were collected from the surface region of the whole shear zone by x-ray CT
(about 10,000 data points). Due to the smaller number of data points,
the scatter of the distributions is larger in these latter cases.
 
As explained before, the center $\Theta_{\rm av}$ of the orientational
distribution was
%
\begin{figure}[ht]
\includegraphics[width=\columnwidth]{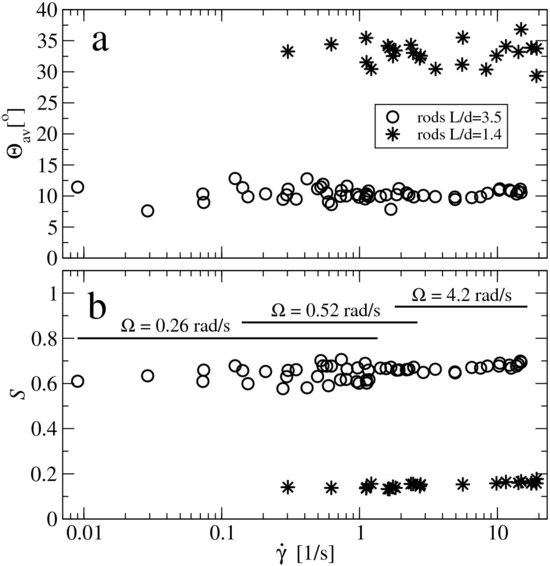}
\caption{(color online). (a) The average orientational
angle $\Theta_{\rm av}$ of the particles with respect to the streamlines and (b)
the order parameter $S$ as a function of the shear rate $\dot{\gamma}$. Data obtained
by optical measurements for glass cylinders with length to width
ratios $L/d= 1.4$ and $3.5$. Both, $\Theta_{\rm av}$
and $S$ are independent of the shear rate.
}
  \label{theta-gamma}
\end{figure}
determined by fitting a Gaussian, and the order parameter $S$, defined as the largest
eigenvalue of \T, was calculated. Note that $S$ is related to the width 
of the angular distribution $\sigma$. As seen in Fig.~\ref{theta-gamma},
$\Theta_{\rm av}$ and $S$ (as well as $\sigma$, not displayed)
are practically shear rate independent across more than three decades of $\dot{\gamma}$.
The slight decrease of $S$ observed for two samples of rice \cite{bosz2012}
is not observed here for cylinders.
The data points were collected using three different rotation rates
($\Omega=$ 0.26, 0.52 and 4.2 rad/s) of the bottom plate.
The spatial variation of $\dot{\gamma}$ in the shear zone allows to span a
relatively wide range of $\dot{\gamma}$ for each rotation rate, as 
marked in  Fig.~\ref{theta-gamma}(b).
In terms of inertial number (defined as $I=\dot{\gamma}d\sqrt{\rho/P}$, where
$\rho$ is the density of particles and for the hydrostatic pressure $P$ 
we use an indicative value taken at depth $d$), our measurements
span the range between 0.0002 to 0.4.

The fact that the average orientation of the particles in the aligned
state is shear rate independent is similar to the well studied \textit{flow
alignment} of nematic liquid crystals. We have quantitatively compared our
observations to the properties of nematics in \cite{bosz2012}, here we only
wish to stress that the strong similarities with nematics indicate that the
shear alignment is robust and is of geometrical origin. Namely, even if the
aspect ratios of the building blocks (molecules and granulates) are similar,
there are other important differences between the two systems, e.g., thermal
fluctuations are important for molecules while they are negligible in the
granular case. Also, the interparticle forces have very different character:
the molecules experience attractive forces (dipole-dipole,
van der Waals), while the granular particles investigated here interact only
via hard core repulsion and friction.

\begin{figure}[ht]
\includegraphics[width=\columnwidth]{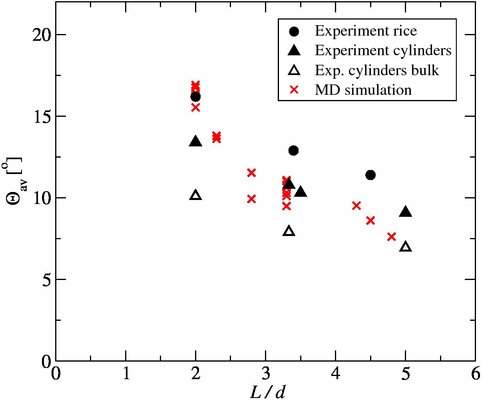}
\caption{(color online). The average orientational angle
$\Theta_{\rm av}$ as a function of the length to diameter ratio $L/d$ of the particles
obtained by experiments ($\bullet$,$\blacktriangle$,$\vartriangle$)
and numerical simulations ($\times$).
}
  \label{theta-s}
\end{figure}

As seen in Fig.~\ref{theta-gamma}, the average alignment angle
$\Theta_{\rm av}$  depends only on the properties
of the particles. The shear rate independent $\Theta_{\rm av}$ has been
determined for all eight samples and it is presented as a function of the aspect
ratio $L/d$ in Fig.~\ref{theta-s}.
The plot combines the bulk and surface data from x-ray tomography, the surface data
from optical experiments, and the numerical results in bulk plane Couette flow.
We find a systematic decrease of $\Theta_{\rm av}$ with
increasing length to width ratio for a given shape family (rice or cylinders).
Results of the numerical simulations nicely agree with the experimental data.
%
At comparable aspect ratios, the alignment angle appears to be systematically smaller for
cylinders than for ellipsoids.

%
\begin{figure}[ht]
\includegraphics[width=\columnwidth]{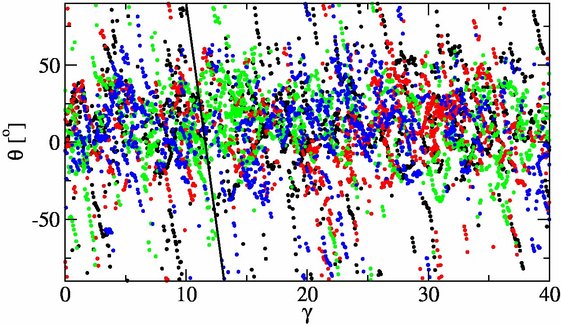}
\caption{
(color online). Evolution of the particle orientations $\theta$ as a function of
strain $\gamma$ in the stationary state for glass cylinders with
$L/d=3.5$. Data obtained by optical particle detection. The black line corresponds
to $d\theta/d\gamma=-1$ (in radians). For better visualization of the trajectories 
we used 4 different colors.
}
  \label{trajectories}
\end{figure}

Focusing on the dynamics of individual grains during stationary
shear, the orientation of the majority of the particles stays within a range of
$\Theta_{\rm av}-\sigma < \theta <\Theta_{\rm av}+\sigma$, where $\sigma$ is the
standard deviation of the Gaussian fit of the orientational distributions.
This range is seen as a dense band in Fig.~\ref{trajectories} between the angles
of $-20^\circ$ and $35^\circ$ for the case of glass cylinders with $L/d = 3.5$.
In this range the particles rotate forward or
backward with essentially equal probability. Outside this range the particles
are rarely seen and they rotate relatively quickly forward as it is indicated by the
steep trajectories in Fig.~\ref{trajectories}.
For these measurements data have been taken from the center of
the shear zone, where the shear rate $\dot{\gamma}$ is nearly constant.

%
\begin{figure}[ht]
\includegraphics[width=\columnwidth]{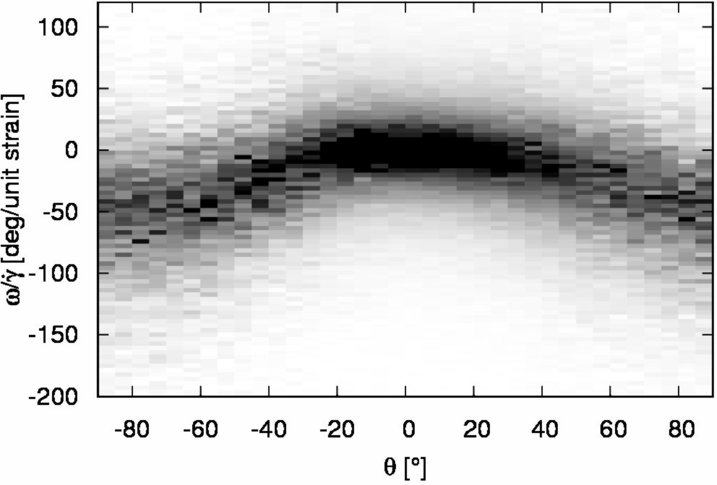}
\caption{
Distribution of the rotation velocity $\omega$ of the particles
as a function of their orientation $\theta$ in the stationary shear aligned
state for glass cylinders with $L/d=3.5$. Data obtained by optical particle detection.
In the grayscale plot black corresponds to the highest probability density.
}
  \label{omegadist}
\end{figure}

The distribution of the particle rotation velocity $\omega={\rm d}\theta/{\rm d}t$ 
has been determined
as a function of $\theta$ and the resulting plot is shown in Fig.~\ref{omegadist}.
Here $\omega$ is normalized by the average shear rate at the location of the
particle. As it is seen $\omega/\dot{\gamma}$ is better defined for the above
mentioned range ($\Theta_{\rm av}-\sigma < \theta <\Theta_{\rm av}+\sigma$) near
$\Theta_{\rm av}$, while the distribution is wider (i.e.  $\omega$ is less well
defined) near $\theta=\Theta_{\rm av}+\pi/2$.  At $\theta=\Theta_{\rm av}+\pi/2$
we find an average rotation speed $\omega/\dot\gamma\approx -1$ rad/unit strain.
Thus, at this angle the rotation speed of the particles is maximal, it reaches 
the value corresponding to the local vorticity of the shear flow.

We calculated the averages of the rotation velocity distributions
normalized by the shear rate, $\omega_{\text{av}}/\dot{\gamma}$, for various
materials, and plotted them against $\theta$.
%
\begin{figure}[ht]
\includegraphics[width=\columnwidth]{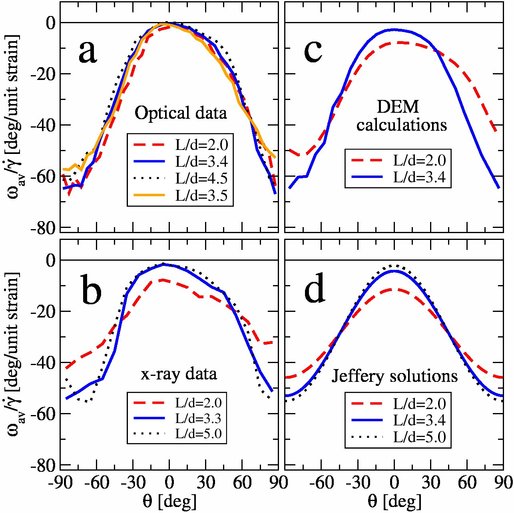}
\caption{
(color online). Normalized average rotation velocity of the particles
$\omega_{\text{av}}/\dot{\gamma}$
as a function of their orientation $\theta$
for (a) rice and glass cylinders obtained by optical measurements
(b) pegs obtained by x-ray tomography, (c) numerical simulations
and (d) hard ellipsoids in a sheared liquid as calculated by Jeffery \cite{je1922}.
}
  \label{omegalperd}
\end{figure}
Data from optical measurements are presented in Fig.~\ref{omegalperd}(a)
for rice grains with $L/d= 2.0$, $3.4$, $4.5$, and glass cylinders with $3.5$, while
Fig.~\ref{omegalperd}(b) shows the results obtained by x-ray tomography
for pegs with $L/d= 2.0$, $3.3$ and $5.0$. Figure~\ref{omegalperd}(c) presents
results obtained by numerical simulation (DEM). We have also included the result
of the calculation by Jeffery \cite{je1922}, which describes the rotation velocity
of a hard prolate ellipsoid immersed in a sheared viscous liquid. All four panels of
Fig.~\ref{omegalperd} show similar tendencies, viz. the particles rotate faster
when their orientation is far from the preferred one and rotate slower when they
are closer to it. The ratio of this fast and slow rotation speeds appears to be smaller
for shorter grains. In other words, longer grains show a stronger angular dependence of the 
rotation rates. 
Differences between our results for the granular particles and the Jeffery solutions
arise from the grain-grain interactions.

\subsection{Transients into the aligned steady state}

\begin{figure}[ht]
\includegraphics[width=\columnwidth]{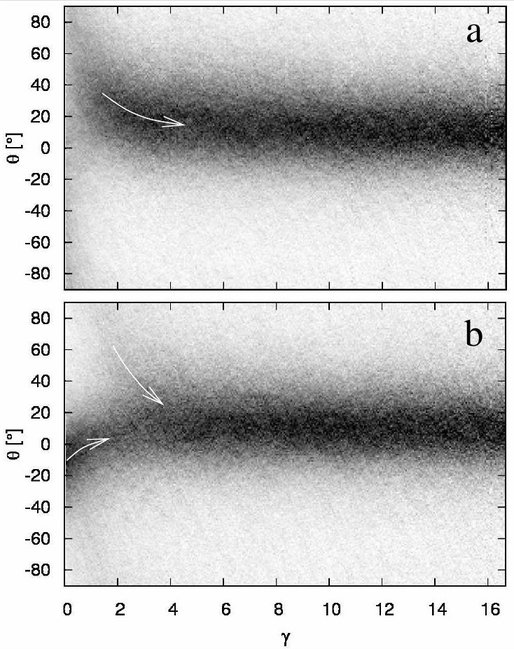}
\caption{
Evolution of the orientational distribution of the particles
(a) starting from an initially random orientation (b) from an aligned 
state after shear reversal.
Both measurements were performed with glass cylinders with $L/d=3.5$, 
data obtained by optical particle detection.
In the grayscale plot, black corresponds to the highest probability density.
The white arrow indicates the evolution of the center of the distribution for
case (a).
For case (b) the initially ordered state appears to split into two independently
relaxing distributions, marked by white arrows.
}
  \label{angle-evolution}
\end{figure}

So far, we have discussed the properties of the stationary state. In the following,
we focus on the dynamics observed in the transient during which the system reaches
the above described stationary state. Two types of experiments have been carried
out: (a) starting from an initially random configuration, and (b) reversing the
shear direction, i.e., starting from an oppositely shear aligned state.
The evolution of the distribution of particle angles has been investigated for
glass cylinders with $L/d=3.5$ taking (a) 1350 and (b) 1700 independent
measurements, respectively. These measurements were performed using surface optical
imaging, and the resulting probability distributions are plotted on
Fig.~\ref{angle-evolution} as a function of the strain (similarly to Fig.~\ref{trajectories}).

%

\begin{figure}[ht]
\includegraphics[width=\columnwidth]{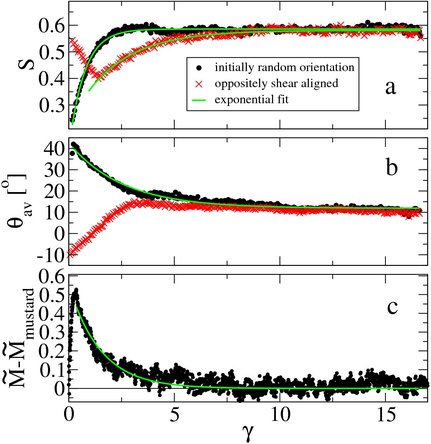}
\caption{
(color online). Evolution of the (a) order parameter and the (b) average alignment angle during both
type of transients. Same measurement as in Fig.~\ref{angle-evolution}.
(c) Evolution of the applied torque needed to maintain stationary rotation
of the bottom plate starting from an initially random orientation.
All measurements {are} for glass cylinders with $L/d=3.5$. Continous lines are fits
assuming exponential relaxation, see text for the numerical values of the
exponents.
}
  \label{transient-curve}
\end{figure}

For the initially random system (a) we find a continuously increasing order and
a decreasing alignment angle [see white arrow in Fig.~\ref{angle-evolution}(a)]
both converging towards their stationary values.
The evolution of the order parameter $S$ and that of the average angle
$\theta_{\rm av}$ are shown as functions of $\gamma$ in Figs.
\ref{transient-curve}(a)-(b). As it is seen, ordering develops faster than
the convergence of the shear alignment angle. This can be quantified by
the characteristic strain parameters of the approximately exponential 
relaxation, which are $\gamma_S=0.73$ and $\gamma_\theta=2.57$, respectively.

For initial condition (b) we find that the average angle rotates in the
\textit{opposite} direction, from $-\Theta_{\rm av}$ to $\Theta_{\rm av}$.
i.e. the majority of the grains turns backwards and only a smaller
fraction rotates forward as indicated by the two white arrows in
Fig.~\ref{angle-evolution}(b).
During this procedure the order parameter initially decreases then increases
back to its stationary value as seen in Fig.~\ref{transient-curve}(a).
Interestingly the strain which is necessary to reach the stationary angle is about
the same for the two initial conditions, but the strain needed to reach
the stationary value of the order parameter is significantly different.
To destroy and then recover the order takes about three times as much strain
($\gamma_S=2.19$) compared to alignment from an initially random state ($\gamma_S=0.73$).

%

\begin{figure}[ht]
\includegraphics[width=\columnwidth]{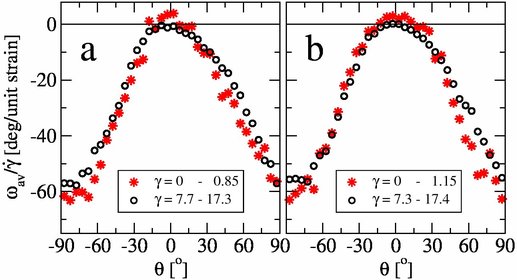}
\caption{
(color online). Normalized average rotation velocity of the particles
$\omega_{\text{av}}/\dot{\gamma}$
as a function of their orientation $\theta$
during the two types of transients (a) starting from a random orientation (b) starting from a
reversed flow aligned state. Measurements done for glass cylinders with $L/d=3.5$, 
data obtained by optical particle detection.
}
  \label{omegatransient}
\end{figure}

It is very interesting to compare the normalized average particle rotation
speed $\omega_{\text{av}}/\dot{\gamma}$ as a function of $\theta$ for these two
different transients and the steady state, see Figs.~\ref{omegatransient}(a)-(b).
In the stationary state, grains in all orientations $\theta$ rotate on average 
with negative $\omega$ (cf. Fig.~\ref{omegatransient} black circles).
During the transients, we find a positive mean rotation velocity for grains between 
$-15^\circ$ and $15^\circ$.
This means, that for both cases (a) and (b) these particles rotate against
the direction dictated by the shear flow, while the rest of the
particles rotate forward.
Here, for the transients, the difference between cases (a) and (b) is that for the
\textit{``oppositely shear aligned''} transient the number of backward rotating grains
is much larger due to the nature of the initial configuration. This is why the
average reorientation of the grains $\theta_{\rm av}(\gamma)$ differs in the two cases 
(see white arrows indicating the grain rotation direction in Fig.~\ref{angle-evolution}
and the curves in \ref{transient-curve}(b)).

%

Additional measurements were performed for the initially random configuration
(a). In these measurements we aimed to monitor the evolution of the effective friction
of the sample during the transient. For this purpose we measured the torque $M$ needed 
for a stationary rotation of the bottom plate, which was rescaled by the asymptotic 
value $M_\infty$ as $\widetilde M=M/M_\infty$. A reference dataset was also recorded 
for nearly spherical mustard seeds, which is used to correct for effects not directly 
related to the reoriantation of the sample.
The rescaled and corrected curve is shown in Fig.~\ref{transient-curve}(c).
The first thing to note is that the ratio of the initial and final values of $\widetilde{M}$
is around 1.5, suggesting that the effective friction of the unoriented sample
is about $50\%$ higher than that of the shear aligned one.
The second observation is, that the evolution of the effective friction of the material
can be described by an exponential decay of the excess torque as
$\widetilde{M}  -\widetilde{M}_{\text{mustard}} \approx (\widetilde{M}_{\rm max}-1)
e^{-\gamma/\gamma_{\text{M}}}$.
This exponential fit results $\gamma_{\text{M}}=1.46$ which is in between the
characteristic values obtained for $\gamma_{\text{S}}$ and $\gamma_\theta$.
Altough the experimental curve is fairly noisy, one can pursue a more detailed
analysis by fitting the data with the sum of two exponentials with decay rates of
$\gamma_{\text{S}}$ and $\gamma_\theta$. The resulting amplitudes suggest that
the evolution of $S$ has a stronger influence on the evolution of $M$ compared to
the contribution of $\theta_{\rm av}$.

\section{Conclusions}
\label{conclusion}

The orientation and rotation of elongated granular particles has been tracked 
in shear flow. In the asymptotic state, the time and ensemble
averaged orientation of the particles forms a small angle $\Theta_{\rm av}$ with
the streamlines.
This shear alignment angle is independent of shear rate and decreases with
increasing particle aspect ratio.
The characteristics of the asymptotic alignment determined from optical experiments in this
study are in good agreement with experiments obtained from x-ray computed 
tomography \cite{Wegner} of cylindrical pegs.  

The average rotation speed of individual particles as function of the
orientation $\omega_{\rm av}(\theta)$ shows that particles rotate faster if their
orientation is perpendicular to the preferred average orientation and their
rotation is slower when their orientation is near $\Theta_{\rm av}$.
The ratio of this fast/slow rotation increases with grain aspect ratio,
similarly to the case of a rotating hard ellipsoid in a sheared liquid as
calculated by Jeffery \cite{je1922}.
In the stationary shear aligned state the net rotation rate of particles is always negative
i.e. the sense of rotation is dictated by the shear flow.
Transients into the asymptotic alignment have been studied starting from an (a) 
initially random configuration and (b) an oppositely shear aligned state. 
The character of the transients depends upon the initial order and alignment.
The transient is longer for case (b) where initial order is partially destroyed 
and then recovered again. A common feature of the two transients is that the particles
with orientation just beyond $\Theta_{\rm av}$ rotate backwards, and reach the
preferred angle on a shorter way.

The evolution of the mechanical resistance of the sample measured during the transient 
indicates a $30\%$ reduction of the materials effective
friction during the course of the shear induced ordering process.


\begin{acknowledgments}
The authors are thankful for G. Rose, chair of the department for
Healthcare Telematics and Medical Engineering of the
Otto von Guericke University, Magdeburg, and appreciate
technical help by G. T\"or\"os.
Financial support by DAAD-M\"OB researcher exchange program (grant No. 29480) is acknowledged.
J.T. acknowledges the support of the German Research Foundation (DFG grant BR 3729/1).
\end{acknowledgments}

\end{document}